\documentclass[reprint, amsmath,amssymb,aps,twocolumn, prb,superscriptaddress]{revtex4-1}
\usepackage{amsfonts,amssymb}
\usepackage{amsmath}
\usepackage[none]{hyphenat}
\usepackage{graphicx}
\usepackage{bm}          
\usepackage{natbib}
\usepackage{soul} 
\usepackage{color}
\usepackage{longtable}
\usepackage{ textcomp }
\usepackage{verbatim}
\usepackage[pdftex,
            pdfauthor={Konstantin V. Reich},
            pdftitle={Dielectric Constant and Charging Energy in Array of NCs},
            pdfkeywords={nanocrystals, transport},
            pdfproducer={Latex with hyperref},
            pdfcreator={pdflatex}]{hyperref}
\begin{document}


\title{Dielectric Constant and Charging Energy in Array of Touching Nanocrystals}
\author{K. V. Reich}
\email{kreich@umn.edu}
\affiliation{Fine Theoretical Physics Institute, University of Minnesota, Minneapolis, MN 55455, USA}
\affiliation{Ioffe Institute, St Petersburg, 194021, Russia}
\author{B. I. Shklovskii}
\affiliation{Fine Theoretical Physics Institute, University of Minnesota, Minneapolis, MN 55455, USA}
\date{\today}

\begin{abstract}
We calculate the effective macroscopic dielectric constant  $\varepsilon_a$ of a periodic array of spherical nanocrystals (NCs) with  dielectric constant $\varepsilon$ immersed in the medium with dielectric  constant $\varepsilon_m \ll \varepsilon$. For an array of NCs with the  diameter $d$ and the distance $D$ between their centers, which are  separated  by the small distance $s=D-d \ll d$ or touch each other by  small facets with radius $\rho\ll d$ what is equivalent to $s < 0$, $|s|  \ll d$ we derive two analytical asymptotics  of the function  $\varepsilon_a(s)$ in the limit $\varepsilon/\varepsilon_m \gg 1$. Using  the scaling hypothesis we interpolate between them near $s=0$ to obtain  new approximated function $\varepsilon_a(s)$ for  $\varepsilon/\varepsilon_m \gg 1$. It agrees with existing numerical  calculations for $\varepsilon/\varepsilon_m =30$, while the standard  mean-field Maxwell-Garnett and Bruggeman approximations fail to  describe percolation-like behavior of $\varepsilon(s)$ near $s = 0$.  We  also show that in this case the charging energy $E_c$ of a single NC in  an array of touching NCs has a non-trivial relationship to  $\varepsilon_a $, namely $E_c = \alpha e^2/\varepsilon_a d$, where $\alpha$ varies from 1.59 to 1.95 depending on the  studied three-dimensional  lattices. Our approximation for $\varepsilon(s)$ can be used instead of  mean field Maxwell-Garnett and Bruggeman approximations to describe  percolation like transitions near $s=0$ for other material  characteristics of NC arrays, such as conductivity.
\end{abstract}
\date{\today}

\maketitle

\begin{figure}
 \includegraphics[width=0.45\textwidth]{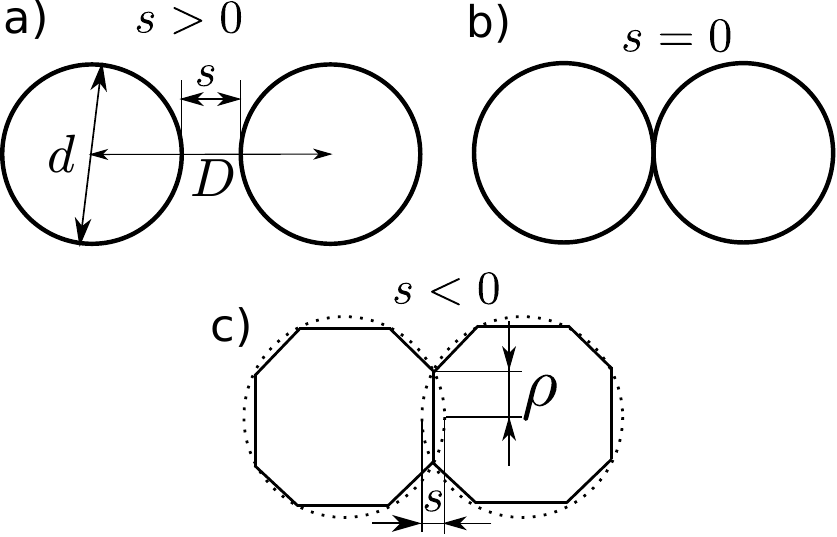}\\
 \caption{Cross-sections of two  spherical NCs with diameter $d$ and distance $D$ between centers. (a) NCs do not touch each other ($s=D-d > 0$).  (b) Two spherical NCs touch in one point ($s=0$). (c) NCs touch each other by disk-like facets with the radius $\rho$. One can say that for this case $s < 0$ (see below). }\label{fig:three_situations}
 \end{figure}

Semiconductor nanocrystals (NCs) can be used as building blocks for new solid materials with bulk properties, which do not exist in conventional solids. From 
almost monodisperse spherical NCs with a few nanometer diameter which have good and size-tunable optical properties one can assemble closely packed NC arrays with a three-dimensional (3D) periodic structure ~\cite{Murray-AnnuRev30-2000,Kagan_QD_review}. 
Spacing between NCs $s$ which usually is much smaller than the NC diameter $d$ may be determined by passivating ligands and tuned by ligand's length. In arrays of bare NCs they can touch in one point or by small facets. Fig. \ref{fig:three_situations} illustrates all three cases. 

For device applications such as light emitting diodes, photovoltaics or transistors NC arrays have to be conducting. One can introduce electrons via doping semiconductor NCs by donors. At concentrations of electrons below the critical concentration of the metal-insulator transition, electrons are localized in each NC and the conductance is due to the variable range hopping  of electrons between NCs. Still at zero temperature such a NC array is an insulator~\cite{Ting_MIT} characterized by a real macroscopic dielectric constant $\varepsilon_a$. Its magnitude determines the characteristic temperature of the Efros-Shklovskii variable range hopping~\cite{Ting_Si,localization_length_NC}. 

To facilitate the electron hopping transport without loss of optical performance related to the spherical shape, NC arrays are made from NCs which touch each other in a point or via small facets \cite{facet_PbSE_mobility_1,facet_PbSE_mobility_2,transport_lattice_QD}. This paper is concerned with calculations of the macroscopic dielectric constant  $\varepsilon_a$ of such arrays. We consider a periodic array with lattice constant $D$  of spherical NCs with diameter $d$ made from a semiconductor with a large dielectric constant $\varepsilon$, which are embedded in an insulating medium with dielectric constant $\varepsilon_m \ll \varepsilon$. Many NC arrays studied in literature have large ratio $\varepsilon/\varepsilon_m$. This ratio may reach 100 for NCs made from PbSe, PbS or PbTe~\cite{iu_Crawford_Hemminger_Law_2013,Law_ligand_length,Guyot-Sionnest_2012,transport_lattice_QD}. 

We show below that for large $\varepsilon/\varepsilon_m$ the effective dielectric constant $\varepsilon_a$ of the NC array critically depends on a small spacing $s=D-d$ between NCs, which can vanish and change its sign as shown for arrays of different densities in Fig. \ref{fig:three_situations}. This happens because the dielectric constant $\varepsilon_a$ is dominated by small contacts between nearest-neighbor NCs.  We demonstrate that in this case, one has to go beyond the mean-field Maxwell-Garnett and Bruggeman approximations~\cite{review_effective_media_Landauer} and present our own extrapolation formula which much better describes existing numerical results.
We also show below, that the failure of the mean-field  approximations also manifests itself in a non-trivial relationship between $\varepsilon_a$ and the charging energy  $E_c$ of a NC in an array of NCs.
 
In order to calculate $\varepsilon_a$ we imagine that the polarizing electric field has a finite frequency $\omega$, so that we are dealing with NCs with the conductivity $\sigma_0= i\omega \varepsilon/4\pi$ in the medium with the conductivity $\sigma_1=i\omega \varepsilon_m/4\pi \ll \sigma_0$. Then we calculate the imaginary conductance $G(\omega)$ between two nearest-neighbor NCs. Here we concentrate on the simple cubic lattice of NCs and in the end of the paper we present generalization to other lattices. For the simple cubic lattice the conductivity of the whole resistor network  $\sigma_a(\omega) = G(\omega)/d $. This brings us to the real dielectric constant of the NC array

\begin{equation}
\label{eq:dielectric_constant_GG}
\varepsilon_a= 4\pi \frac{G(\omega)}{i\omega d},
\end{equation}
\noindent  in terms of yet unknown conductance $G(\omega)$.

The conductance $G$ between two spheres with conductivity $\sigma_0$ immersed in media with conductivity $\sigma_1 \ll \sigma_0$ with the spacing  $0< s \ll d$ was calculated by Keller~\cite{Keller_1963} in the limit of the infinite ratio $\sigma_0/\sigma_1$. He arrived at $G= (\pi \sigma_{1} /2)\ln (d/2s)$. For our dielectric problem $\sigma_1=i\omega \varepsilon_m/4\pi \ll \sigma_0$ and 

\begin{equation}
\label{eq:conductace_G}
G(\omega) = \frac{i\omega  \varepsilon_m}{8} \ln\left( \frac{d}{2s}\right).
\end{equation}
Using this result we find from Eq. (\ref{eq:dielectric_constant_GG}) that in the limit of infinite $\varepsilon/\varepsilon_m$ 
\begin{equation}
\label{eq:dielectric_constant_diverging}
\varepsilon_a (s)=\frac{\pi}{2}\varepsilon_m \ln\left( \frac{d}{2s} \right).
\end{equation}

\begin{figure}
\includegraphics[width=0.5\textwidth]{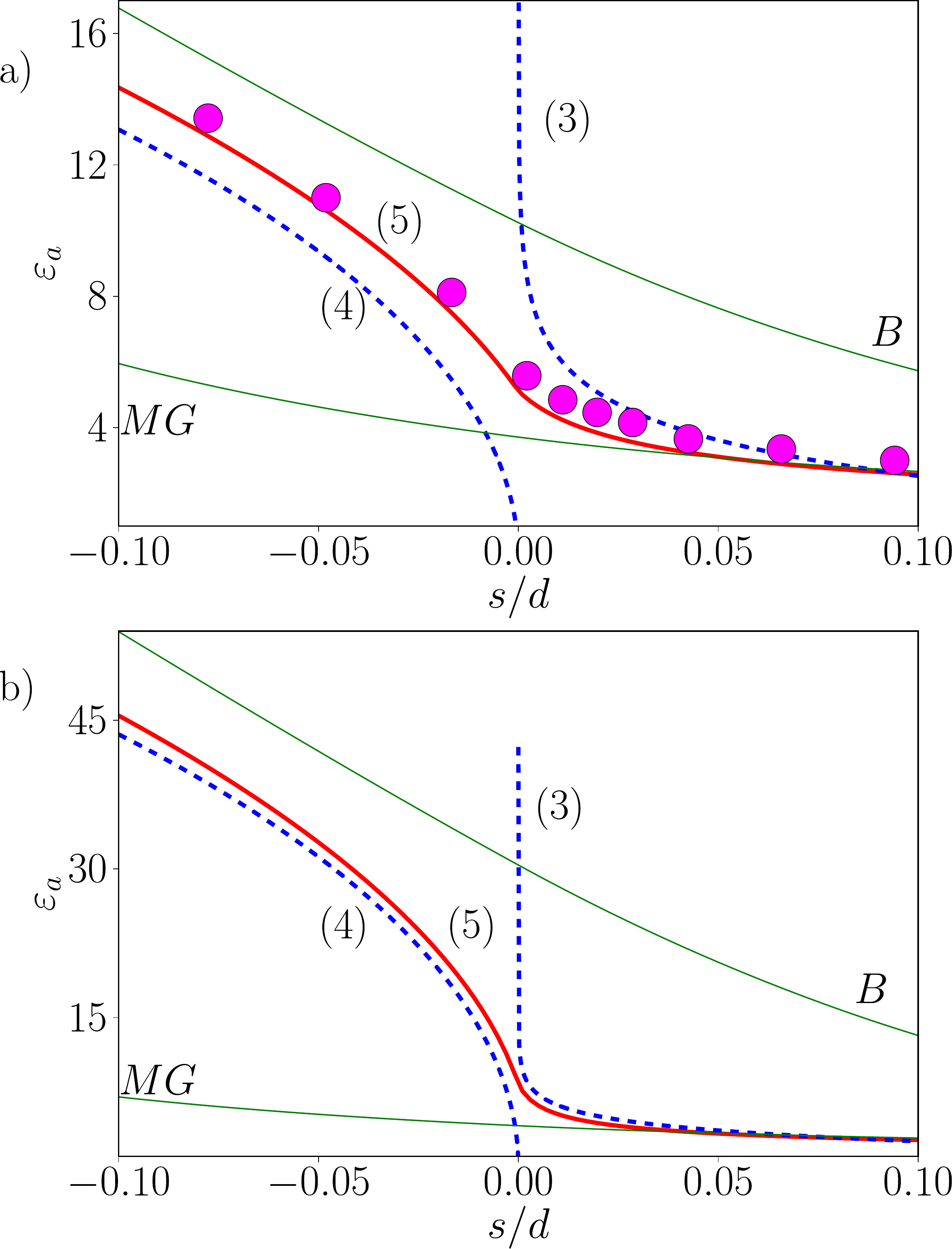}\\
\caption{(Color online) The dependence of the dielectric constant  $\varepsilon_a$ on the spacing $s$ between two spherical NCs for the simple cubic  lattice of NCs in the media with dielectric constant $\varepsilon_m=1$. a) $\varepsilon =30$, b) $\varepsilon =100$. Dashed lines (blue) are our theoretical asymptotics, and the thick solid grey (red) line is our final interpolation expression. Corresponding formulas are given by numbers next to them. The  thin solid grey (green) lines are Maxwell-Garnett (MG) and Bruggeman (B) approximations. Filled circles (magenta) are numerical results from Ref. \onlinecite{dielectric_constant_periodic_numerical_30}.}\label{fig:dielectric_constant_final}
\end{figure}

\noindent We see that $\varepsilon_a (s)$ diverges at $s=0$ as shown in Fig. \ref{fig:dielectric_constant_final} by the dashed (blue) line (\ref{eq:dielectric_constant_diverging}) at $s > 0$. 

Nearest neighbor conductances play the dominant role also for arrays of NCs, which touch by small disc-like facets with radius $\rho$ along (100) axes. Such a geometry can also be viewed as two "intersecting spheres" with diameter $d$ and distance between centers $D$ (see dashed lines in Fig.\ref{fig:three_situations}c). This allows us to introduce $s = D-d < 0$,  express $\rho$  for small $|s|$ as $\rho=\sqrt{|s|d/2}$ and eventually plot $\varepsilon_a (s)$ of faceted NCs in Fig.  \ref{fig:dielectric_constant_final} together with positive $s$ results. 

To calculate the conductance $G$ between two touching by facets NCs we use the known result for the conductance of a circular constriction with radius $\rho$ in a planar insulating diafragm separating two half-space conductors with conductivity $\sigma_0$ each. According to Maxwell \cite{Maxwell_treatise,transport_through_constriction} this conductance is $G= 2\sigma_0\rho$. When $\rho \ll d$ in the first approximation we can consider the contact of two facets as a constriction in the planar diafragm formed by the surrounding media with the small dielectric constant $\varepsilon_m$. Then using $\sigma_0 = i\omega\varepsilon/4\pi$
we find $G(\omega) =2i\omega\varepsilon\rho/4\pi$. Substituting it into Eq. (\ref{eq:dielectric_constant_GG}) 
we arrive at the dielectric constant of the array with $s < 0$ 
\begin{equation}
 \label{eq:constant_touch}
\varepsilon_a(s) = 2 \varepsilon \rho /d =\varepsilon \sqrt{2 \frac{|s|}{d}}.
\end{equation}
This result is plotted at $s < 0$ in Fig.~\ref{fig:dielectric_constant_final} by the dashed (blue) line (\ref{eq:constant_touch}). It depends only on $\varepsilon$ and plays the role of the $s<0$ asymptotics complimentary to the 
$s > 0$ asymptotics Eq. (\ref{eq:dielectric_constant_diverging}). 

In Fig.~\ref{fig:dielectric_constant_final}a we plot the analytical results Eqs. \eqref{eq:constant_touch} and \eqref{eq:dielectric_constant_diverging} together with the result of numerical computation \footnote{\label{footnote_volume_fraction_s} We use the following relationship between $s$ and the volume fraction $f$. For $s>0$ $f=\pi/6 (1+s/d)^{-3}$ and for $s<0$ $f=\pi/6 (1+s/d)^{-3} (1-1.5(s/d)^2(3+s/d))$.} $\varepsilon_a(s)$ in the case $\varepsilon_m=1$, $\varepsilon=30$ from Ref. \onlinecite{dielectric_constant_periodic_numerical_30}. 

One can see that numerical results cross over well between our asymthotic curves Eqs.~(\ref{eq:constant_touch}) and (\ref{eq:dielectric_constant_diverging}) demonstrating a good agreement with our theory. This crossover resembles the one near the percolation threshold in the random mixture of two phases with a very large ratio of conductivities or dielectric constants due to the change of the volume fraction of the phases~\cite{conductivity_near_MIT_ES_1976,Straley_1976_critical_phenomena_resistor_network}.   

Applying the scaling approach of Refs. \onlinecite{conductivity_near_MIT_ES_1976,Straley_1976_critical_phenomena_resistor_network} we can assume that at very large ratio of $\varepsilon/\varepsilon_m$ the crossover between expressions \eqref{eq:dielectric_constant_diverging} and \eqref{eq:constant_touch} happens in the small symmetric critical interval $(-\delta, \delta)$. To find the magnitude of $\delta$ we substitute $\delta$ for $|s|$ into both Eqs.~\eqref{eq:dielectric_constant_diverging} and \eqref{eq:constant_touch} and then equate their right sides. To proceed analytically we use the approximation $\ln x \simeq x^{1/3}$ valid within 10\% for $4 < x <100$. This gives

$$\delta=d \frac{\pi^{6/5}}{2^{11/5}} \left(\frac{\varepsilon_m}{\varepsilon}\right)^{6/5}.$$

Then our final interpolation formula for $\varepsilon_{a}$ is

\begin{equation}
\label{eq:dielectric_constant_cross_final}
\varepsilon_a (s)=
  \begin{cases}
    \dfrac{\pi}{2}\varepsilon_m {\left( \dfrac{d}{2s+2\delta} \right)}^{1/3}    & \quad \text{if } s>0\\
\varepsilon \sqrt{2 \dfrac{|s|+\delta}{d}}      & \quad \text{if } s<0\\
  \end{cases}
\end{equation}
This dependence is shown on Fig.~\ref{fig:dielectric_constant_final} by the full grey (red) line. It agrees quite well with the numerical data.

Eq. \eqref{eq:dielectric_constant_cross_final} can be also used near $s=0$ for averaging  other strongly different material parameters, for example, for the electric  conductivity or the thermal conductivity. 

Results of the Maxwell-Garnett and Bruggeman approximations~\cite{review_effective_media_Landauer} are added to Fig. \ref{fig:dielectric_constant_final} by the solid grey (green) lines marked MG and B correspondingly~\footnotemark[1]. We see that at $\varepsilon = 30$,  $\varepsilon_m = 1$ both mean-field approximations fail to describe the percolation-like transition happening at small $|s|$. The difference between our theory and the two mean-field approximations near $s=0$ is even more pronounced for the case $\varepsilon \geq 100$, important for PbTe, PbS, PbSe applications \cite{iu_Crawford_Hemminger_Law_2013,Law_ligand_length,Guyot-Sionnest_2012}.
(see Fig. \ref{fig:dielectric_constant_final}b).

So far we dealt with the macroscopic dielectric constant $\varepsilon_a(s)$  which describes the response of a NC array to the external electric field on scales much larger than $d$. At the same time, in a NC array with  electron states localized inside each NC a single NC can be charged, say, by an extra electron. Due to the large dielectric constant $\varepsilon$ of the semiconductor, the most of this NC charge is transfered by the dielectric response to the surface of the NC~\cite{Conductivity_Brian_Shklovskii}. As a result, the energy of a charged NC $E_c$ called the charging energy is the unique function of the NC charge $e$ and the NC size $d$
$$
E_c = \frac{e^2}{2C} = \frac{e^2}{\varepsilon_c d},
$$
where $C$ is the capacitance of a NC immersed in the array and the effective dielectric constant $\varepsilon_c = 2C/d$ describes the charging, which is a local response. Therefore, in principle, $\varepsilon_c$ may be different from  $\varepsilon_a$. However, all the NC array literature assumes that $\varepsilon_c = \varepsilon_a$. 

The charging energy $E_c$ can be measured as the activation energy of the nearest neighbor hopping in lightly doped NC arrays, where 
the number of donors per NC is much smaller than unity~\cite{Ting_Si}. In this case all the NCs are neutral in the ground state and the energy $2E_c$ is required for an electron to leave its donor in the doped NC and get transfered to a distant undoped one, because this process creates two charged NCs. Thus, the question whether $\varepsilon_c = \varepsilon_a$ is experimentally verifiable.

For NCs with a relative small dielectric constant $\varepsilon < 10$, when mean-field approximations work well in the whole range of $s$ it is natural to assume that $\varepsilon_c =\varepsilon_a$~\cite{Conductivity_Brian_Shklovskii}. However, when $\varepsilon \geq 30$ and as we saw above the mean-field approaches fail near $|s| \ll d$ we show below that $\varepsilon_c = \varepsilon_a/\alpha $, or 

\begin{equation}
  \label{eq:charging_sc0}
E_c = \alpha\frac{e^2}{\varepsilon_a d},
\end{equation}
where the numerical coefficient $\alpha$ depends on the array structure (see Table I). 

In order to prove Eq.~\eqref{eq:charging_sc0} we start from an important  property of an infinite cubic resistor network made of identical resistors $R$. The resistance between a site of this lattice and infinity is known to be $\beta R$ where $\beta \simeq 0.253$ ~\cite{PhysRevB.70.115317,Lattice_green_function} and $1/\beta$ plays the role of the effective number of parallel resistors $R$ connecting this site to infinity. We show below that $\alpha = 2\pi \beta$. To do that let us return to our cubic NC array and consider it at a finite frequency $\omega$, when for small enough $|s|$  it becomes a cubic resistor network with $R = 1/G(\omega)$. Now the resistance from a site to infinity is $1/i\omega C$. Thus using the above relationship between this resistance and $R$ we get that $i\omega C = G(\omega)/\beta$. Then using Eq. (2) we get

$$
\varepsilon_c = 2C/d = 2 G(\omega)/i\beta\omega d  = \varepsilon_a/(2\pi\beta),
$$
so that for the simple cubic lattice of touching NCs $\alpha = 2\pi\beta$. For this lattice $\beta \simeq 0.253$~\cite{PhysRevB.70.115317,Lattice_green_function}, so that according Eq. (\ref{eq:charging_sc0})

\begin{equation}
  \label{eq:charging_sc}
E_c =1.59 \frac{e^2}{\varepsilon_a d}.
\end{equation}

Below we generalize our results to the body-centered cubic (bcc) and face-centered cubic (fcc) lattices of touching NCs. First we generalize Eq. \eqref{eq:dielectric_constant_GG}. We write the conductivity of a lattice as $\sigma(\omega)=\gamma G(\omega)/d$, where $G(\omega)$ is the conductance  between two nearest-neighbor NCs and the coefficient $\gamma$ is shown in Table I. As a result, the dielectric constant of the NC array is a simple generalization of Eq. \eqref{eq:dielectric_constant_GG}:

\begin{equation}
  \label{eq:dielectric_constant_lattices}
  \varepsilon_a=4\pi \gamma \frac{G(\omega)}{i\omega d}.
\end{equation}

Also the right sides of Eqs. \eqref{eq:dielectric_constant_diverging},~\eqref{eq:constant_touch},~\eqref{eq:dielectric_constant_cross_final} should be multiplied by  $\gamma$. 

Second, we need the effective number  $1/\beta$ of neighboring resistors which spread the current from a NC to infinity for these lattices of NCs. It is known that $\beta = P/Z$, where $Z$ is the nearest neighbor number of the lattice and $P$ is  the inverse probability that a random-walking particle never returns to the origin. The latter is the well known number for all lattices~\cite{Lattice_green_function}. We list coefficients $\beta$, $\gamma$ and $\alpha=2\pi \gamma \beta$ for three NC lattices in Table I.

\begin{table}[t!]
\label{tab1}
  \begin{tabular}{ c | c| c|  c }
    \hline
    Lattice & $\beta$  & $\gamma$  & $\alpha$  \\ \hline
    sc & 0.25 & $1$ & 1.59  \\ 
    bcc & 0.17 &$\sqrt{3}$& 1.84  \\ 
    fcc & 0.11 &$2\sqrt{2}$& 1.95 \\ 
    \hline
  \end{tabular}
 \caption{Parameters for simple(sc), body-centered (bcc) and face-centered (fcc) cubic lattices.}
\end{table}

We see that the nontrivial coefficient $\alpha$ in Eq. (\ref{eq:charging_sc0})
varies in a relatively narrow interval between 1.59 and 1.95. It is likely that $\alpha$ is close to 1.95 for the random dense packing. 

\section*{ACKNOWLEDGEMENT}

The authors would like to thank Han Fu and B. Skinner for helpful discussions. This work was supported primarily by the National Science Foundation through the University of Minnesota MRSEC under Award No. DMR-1420013.

\end{document}